# Wall-induced translation of a rotating particle in viscoelastic fluid

Sai Peng[1,2] and Peng Yu[1,2,3*]

[1]Department of Mechanics and Aerospace Engineering, Southern University of Science and Technology, Shenzhen, 518055, China

[2]Guangdong Provincial Key Laboratory of Turbulence Research and Applications, Southern University of Science and Technology, Shenzhen, 518055, China

[3]Center for Complex Flows and Soft Matter Research, Southern University of Science and Technology, Shenzhen, 518055, China

**Abstract**
Shear-thinning and viscoelasticity are two non-Newtonian fluid properties widely existing in biological fluids. In this study, we found that the translation motion of a rotating particle near a wall speed up firstly, and then slows down with enhancement of fluid viscoelasticity, which is different from the behavior reported in shear thinning fluid (Chen *et al.* J. Fluid Mech. 2021, 927). Our research is carried out by numerical simulation of Navier-Stokes equations combined with Oldroyd-B constitutive model. This work is expected to be helpful to understand the movement of a rotating sphere near a wall in complex fluids comprehensively.

The motion of particles near boundaries has attracted wide attention, because in most real situations, the approach of a boundary is almost inevitable, and it has great biological and technical significance, such as the separation of blood cells and platelets in blood vessels and particle manipulation technology in microfluidic channels [1-3]. In addition, it is necessary to consider the boundary existence of artificial micro-swimmer, which may lead to the enhancement or obstruction of their propulsion and traps or guidance [4-6]. In particular, a kind of artificial micro swimmers called surface walkers or micro-rollers use their interactions with nearby surfaces to generate directional propulsion [7,8]. These micro swimmers are usually driven to rotate by external magnetic field. Boundary symmetry breaking will correct their rotation into translation, just like a wheel rolling on a solid surface. These simple and effective micro-rollers show great opportunities for their applications in targeted therapy and microsurgery.

In Newtonian fluid, the particle-wall interactions at low Reynolds numbers (*Re*) has been well studied [9-12]. Emerging biomedical applications of micromachines in complex biological fluids, however, prompt new questions on the impact of non-Newtonian rheology on these interactions. In the human body fluid environment, the existence of macromolecules, such as protein, DNA molecules, red blood cells, etc., the fluid usually behaves both shear-thinning and viscoelasticity [13]. Weissenberg number (*Wi*) could be used to the ratio of relaxation time to characteristic time of polymer solution. Recently, Chen *et al.* [14] considered fluid's shear-thinning on the speed of wall-induced translation of a rotating particle. In two dimensions, the shear-thinning effect causes a rotating cylinder always to translate in a direction opposite to what might be intuitively expected for an object rolling on the wall. In three dimensions, a rotating sphere may propel either forwards or backwards depending on its rotational frequency and properties of the shear-thinning fluid. However, how is the effect of viscoelasticity rheology on this flow. In this work, through numerical simulation, we report effect of viscoelasticity rheology on the translation-rotation coupling of a particle near a wall. Such coupling is relevant to not only the propulsion of micro rollers but also the near-wall dynamics of swimming bacteria in complex fluids [15-17].

Classical results of wall-induced translation of a rotating object were obtained in a Newtonian fluid in the Stokes regime. For the three-dimensional (3-D) Stokes flow around a rotating sphere near a plane wall, the sphere translates parallel to the wall in a direction consistent with the rolling of a sphere along the wall, without any velocity component normal to the wall, $U = U_x \mathbf{e}_x$; see notation and set-up in Fig. 1(a). The direction of induced translation can be understood as a consequence of the fact that the rotating sphere causes higher velocity gradients in the fluid gap between the sphere and the wall and hence a larger hydrodynamic force on the side of the sphere closer to the wall than that on the other side. The force imbalance thus drives the sphere to translate in a direction expected for a rolling sphere on a solid substrate via friction asymmetry. Here, we revisit these classical results on translation-rotation coupling near a wall and examine how viscoelasticity rheology modifies the coupling in the 3-D case. The calculated spatial domain is shown in Fig. 1(a). Wall-induced translation of a rotating cylinder or sphere of radius a at a distance h above a plane wall. Upon a prescribed rotational velocity $\boldsymbol{\Omega} = -\Omega\mathbf{e}_z$, the particle translates parallel to the wall with an unknown velocity $U = U_x\mathbf{e}_x$.

*Corresponding Author: yup6@sustech.edu.cn

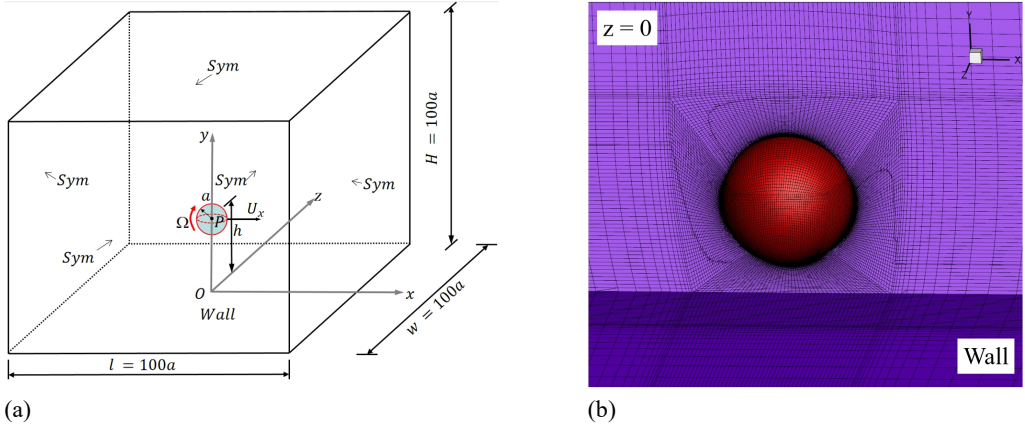

**Fig.1.** (a) The calculated spatial domain; (b) Near sphere mesh distribution view.

In this study, our computational domain is a regular hexagon. The top, front, back, left and right faces are set as symmetrical boundaries to consider a infinite spatial domain problem. The side length of the hexagon is set to $l = H = w = 100a$.

We begun our analysis with the momentum and continuity equations for an incompressible flow in the low-Reynolds-number limit,

$$\rho\mathbf{u}\cdot\nabla\mathbf{u} = \nabla\cdot\boldsymbol{\sigma},\ \nabla\cdot\mathbf{u} = 0, \quad (1)$$

where $\mathbf{u}$ is the fluid velocity, $\rho$ is the fluid's density, $\boldsymbol{\sigma} = -p\mathbf{I} + \boldsymbol{\tau}_s + \boldsymbol{\tau}_p$ is the stress tensor, and $p$, $\boldsymbol{\tau}_s = \eta_s\dot{\gamma}$ ($\dot{\gamma}$ is shear strain rate, $\eta_s$ is the viscosity contribution from the solvent) and $\boldsymbol{\tau}_p$ are the pressure and the shear stress and elastic stress, respectively. To capture the viscoelastic behavior, We use the Oldroyd-B constitutive equation, which is shown to describe effectively the viscosity $\eta$ of different biological fluids [18],

$$(\mathbf{u}\cdot\nabla)\mathbf{c} = (\nabla\mathbf{u})\cdot\mathbf{c} + \mathbf{c}\cdot(\nabla\mathbf{u})^T - (\mathbf{c}-\mathbf{I})/\lambda. \quad (2)$$

There are 10 equations (4 flow equations and 6 conformation tensor equations) in three dimensions. The elasticity satisfies the following constitutive relation,

$$\boldsymbol{\tau}_p = \eta_p\,(\mathbf{c}-\mathbf{I})/\lambda. \quad (3)$$

where $\eta_p$ is the viscosity contribution from the the polymer ($\eta = \eta_s + \eta_p$). Oldroyd-B model is a simple model to describe viscoelasticity is first proposed by James Oldroyd-B [19]. This model holds that the elastic stress and strain rate of polymer solution have a linear relationship similar to Hooke's elasticity. This viscoelastic constitutive model shows the invariant of shear viscosity at various strain rates. Using this model, we can consider the viscoelasticity of fluid separately, and ignore the influence of shear-thinning. Wessienberg number is defined as $Wi = \lambda\Omega$. Reynolds number is defined as $Re = \rho\Omega a^2/\eta$. In this study, it is fixed at 0.01. The polymer viscosity ratio at vanishing shear rate is defined as $\beta = \eta_s/\eta$, which is a measurement of polymer concentration and molecular characteristics. Fluid's elasticity is positively correlated with $Wi$ and $1-\beta$.

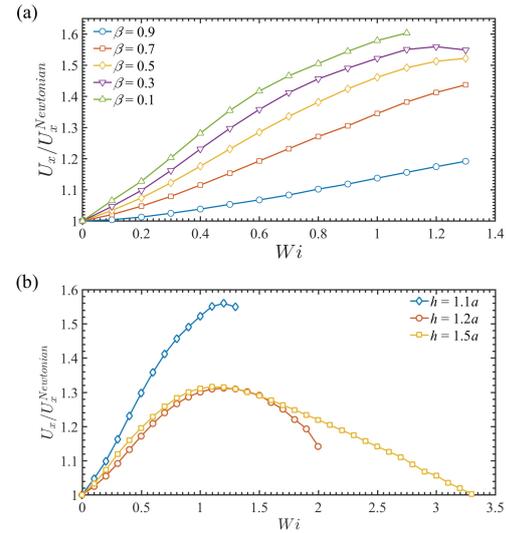

**Fig.2.** Wall induced velocity speed for (a) $h = 1.1a$, (b) $\beta = 0.3$ in viscoelastic fluids. The speed is normalized by that in Newtonian fluid.

To solve the above equations numerically, a finite volume commercial solver ANSYS FLUENT 15.0 associated with a self-developed user defined function (UDF) for the elastic stress transport equation, i.e., Eq. (2), is applied [20-22]. This method has been proved to be effective method to solve the flow problems related to viscoelastic fluid. The QUICK scheme is adopted to discretize the convection terms of both momentum conservation equation and transport equation of configuration tensor. All boundary conditions of tensor $\mathbf{c}$ are approximate no flux. In order to avoid the mesh transformation when the sphere moves, the non-inertial frame of reference method is adopted. In this calculation, the speed of the sphere is assumed stable. For the reference method, an opposite speed $-U_x$ is loaded on the wall. The value of $U_x$ is obtained by trial-and-error method so that the drag of the cylinder is around zero. We validated the numerical method against solution of Chen *et al.* [14] in a Newtonian fluid in Supplement A. The mesh number is of the order of 1~2 × $10^6$ for the 3-D simulations, depending on the distance of the cylinder/sphere from the wall. The simulation accuracy validation of viscoelastic fluid is given in Supplement B.

In our simulation, we focus $Wi$, $\beta$ and $h$ on the transitional speed of the sphere. First, we fixed $h$ at 1.1$a$ to study $Wi$ and $\beta$ effects. As shown in Fig. 2(a), the dimensionless speed ($U_x/U_x^{Newtonian}$) is enhanced in viscoelastic fluids for $\beta$ (from 0.1 to 0.9) when $Wi$ is less than a critical number $Wi_c$. Smaller $\beta$, a larger $U_x/U_x^{Newtonian}$. The greater the elasticity of the fluid, the greater the increase in the velocity of the sphere. However, when $Wi$ is over $Wi_c$, $U_x/U_x^{Newtonian}$ goes a reversed trend. For example, for $\beta$ = 0.3, the speed is getting smaller when $Wi$ is over 1.2. It could be guessed that the corresponding $Wi_c$ also exists for other $\beta$, and when $Wi$ exceeds $Wi_c$, the speed will slow down. However, due to the instability of numerical calculation, our calculation can't be performed at higher $Wi$. Second, we fixed $\beta$ at 0.3 to study the influence of $Wi$ and $h$. For another $h$, such as $h$ = 1.2$a$ and 1.5$a$. The speed can also be enhanced in viscoelastic fluids when $Wi$ is less than a critical number $Wi_c$. When $Wi$ is over $Wi_c$, the speed will then be slowed. For $h$ = 1.5$a$, we could calculate up to $Wi$ = 3.3. At this $Wi$, the speed is even lower than that in Newtonian fluid.

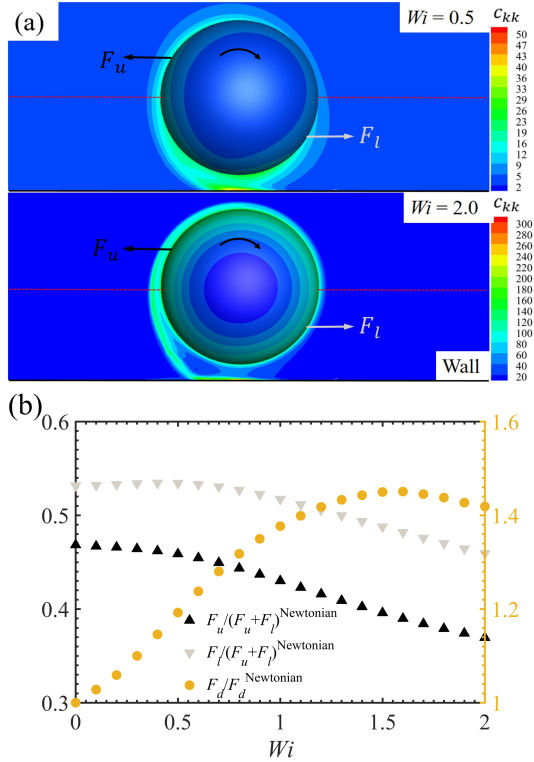

**Fig. 3.** (a) Distribution of trace of tensor configuration for $Wi$ = 0.5 and $Wi$ = 2.0; (b) The hydrodynamic force on the rotating sphere. The sphere is not free to translate. $\beta$ is fixed at 0.3 and $h$ = 1.2$a$.

In viscoelastic flow, the existence of elasticity plays a crucial role. To develop some understanding of these results, we examine the trace of conformation tensor distribution around the rotating sphere, when it is not free to translate. Under this set-up we analyze the hydrodynamic force on the rotating sphere and probe any resulting force imbalance that drives a translation. Due to the velocity gradient of the wall, the elasticity is concentrated on the sphere. The existence of elastic stress inhibits the rotating effect of the sphere, ie., produces an opposite torque, that is, an opposite force to $F_l$ and $F_u$. As shown in Fig. 3(b), the force of both $F_l$ and $F_u$ decrease in viscoelastic fluid. The behaviour of $F_l$ and $F_u$ in viscoelastic fluid is similar to that in shear-thinning fluid, both of which decrease with the increase of rotating rate.

However, the presence of the wall breaks the up-down symmetry and causes higher velocity gradients in the fluid gap below the cylinder than in the fluid above it. The elastic stress developed from the gap to upstream of the sphere, then above the sphere, finally downstream of the sphere along the rotating main streamline. At low $Wi$, such as $Wi$ = 0.5, the maximum elastic stress distributes in the gap, then upstream of the cylinder. The more concentrated the elasticity, the higher the pressure distribution. Therefore, $F_l$ decrease less than $F_u$ at low $Wi$, as shown in Fig.3(b), which results in the total force $F_d = F_l - F_u$ increase when $Wi$ less than 1.6. At high $Wi$, such as $Wi$ = 2.0, the elastic stress distributes uniformity in the circumferential direction, as shown at bottom panel in Fig. 3(a). Due to the increase of elastic stress in the back of the sphere, $F_u$ is reduced at this stage accelerate. The total force $F_d = F_l - F_u$ decrease when $Wi$ is over 1.6. Due to the force on the sphere increases first (≤1.6) and then decreases (>1.6), when it is not free to translate, the translation velocity $U_x$ increase first then decrease as $Wi$ increases.

In this letter, through numerical simulation of Navier Stokes equations combined with Oldroyd-B model, the translational motion of a rotating particle near a wall in viscoelastic fluid is studied. It is found that the translation motion speeds up firstly, and then slow down with the enhancement of fluid viscoelasticity. However, Chen *et al.* [14] reported in shear thinning fluids, the translational speed of sphere becomes lower, even change direction. Viscoleasticity is found be different from the behavior reported in shear-thinning fluid. Combining our work with Chen *et al.*, it may fully understand the influence of fluid's complex rheology on wall-induced translation of a rotating particle.

The author P Yu would like to thank the financial support from Shenzhen Science and Technology Innovation Commission (Grant No. JCYJ20180504165704491), Guangdong Provincial Key Laboratory of Turbulence Research and Applications (Grant No. 2019B21203001), the National Natural Science Foundation of China (NSFC, Grant No. 12172163, 12002148, 11672124). This work is supported by Center for Computational Science and Engineering of Southern University of Science and Technology.

**Reference**

1. L. Crowl and A. L. Fogelson, J Fluid Mech **676**, 348 (2011).
2. J. Zhang, S. Yan, D. Yuan, G. Alici, N. T. Nguyen, M. E. Warkiani, and W. H. Li, Lab Chip **16**, 10 (2016).
3. A. Duzgun and J. V. Selinger, Phys Rev E **97** (2018).
4. D. Takagi, J. Palacci, A. B. Braunschweig, M. J. Shelley, and J. Zhang, Soft Matter **10**, 1784 (2014).
5. J. Elgeti and G. Gompper, Eur Phys J-Spec Top **225**, 2333 (2016).
6. S. E. Spagnolie and E. Lauga, J Fluid Mech **700**, 105 (2012).
7. Y. Alapan, U. Bozuyuk, P. Erkoc, A. C. Karacakol, and M. Sitti, Sci Robot **5** (2020).
8. D. Ahmed, A. Sukhov, D. Hauri, D. Rodrigue, G. Maranta, J. Harting, and B. J. Nelson, Nat Mach Intell **3** (2021).
9. W.R. Dean and M.E. O'neill, Mathematika **10** (1963).
10. L. G. Leal, Annu Rev Fluid Mech **12**, 435 (1980).
11. S. H. Lee and L. G. Leal, J Fluid Mech **98**, 193 (1980).
12. S. H. Lee, R. S. Chadwick, and L. G. Leal, J Fluid Mech **93**, 705 (1979).
13. M. Brust, C. Schaefer, R. Doerr, L. Pan, M. Garcia, P. E. Arratia, and C. Wagner, Phys Rev Lett **110** (2013).
14. Y. Chen, E. Demir, W. Gao, Y. N. Young, and O. S. Pak, J Fluid Mech **927** (2021).
15. Z. J. Qu and K. S. Breuer, Phys Rev Fluids **5** (2020).
16. E. Lauga, Phys Fluids **26** (2014).
17. E. Lauga, W. R. DiLuzio, G. M. Whitesides, and H. A. Stone, Biophys J **90**, 400 (2006).
18. H. A. C. Sanchez, M. R. Jovanovic, S. Kumar, A. Morozov, V. Shankar, G. Subramanian, and H. J. Wilson, J Non-Newton Fluid **302** (2022).
19. J. G. Oldroyd, Proc R Soc Lon Ser-A **200**, 523 (1950).
20. S. Peng, Y. L. Xiong, X. Y. Xu, and P. Yu, Phys Fluids **32** (2020).
21. Y. L. Xiong, S. Peng, M. Q. Zhang, and D. Yang, J Non-Newton Fluid **272** (2019).
22. Y. L. Xiong, S. Peng, D. Yang, J. Duan, and L. M. Wang, Phys Fluids **30** (2018).
23. M. A. Alves, F. T. Pinho, and P. J. Oliveira, J Non-Newton Fluid **97**, 207 (2001).
24. M. A. Hulsen, R. Fattal, and R. Kupferman, J Non-Newton Fluid 127, 27 (2005).
25. Y. R. Fan, R. I. Tanner, and N. Phan-Thien, J Non-Newton Fluid 84, 233 (1999).

**Supplement**

**A. Comparison of wall-induced translation of a rotating particle in Newtonian fluid**

We compare the wall-induced translation speed of a rotating sphere in Newtonian fluid as list in table 1. Our numerical results coincide well with Chen *et al.* [14].

**Tab. S1.** Wall-induced translation speed of a rotating particle in Newtonian fluid.

| $h$ | $1.1a$ | $1.5a$ | $2.0a$ |
|---|---|---|---|
| Present | 0.05236 | 0.013677 | 0.004996 |
| Chen *et al.* | 0.05254 | 0.013757 | 0.004762 |

**B. Viscoelastic flow validation**

To validate the present numerical method, a series of simulations based on the Oldroyd-B model are performed for the flow past cylinder in a channel with $BR = 50\%$ and the numerical results are compared with published data. In these simulations, the Reynolds number is set as zero, i.e., the convection terms in the momentum equations are ignored. In our simulation, $Re$ is set as 0.01. The viscosity ratio $\beta$ is fixed at 0.59 while $Wi$ is varied from 0 to 0.8. As shown in Fig. S1, our results coincide well with the results of Hulsen *et al.*, Fan *et al.* and Alves *et al.* [23-25].

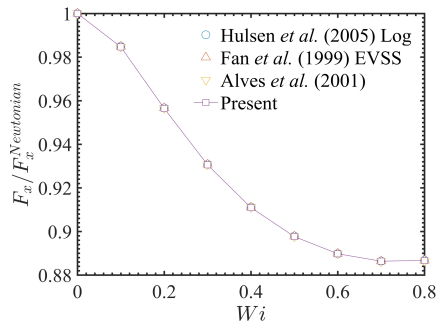

**Fig. S1**. The hydrodynamic force on a confined circular cylinder. $\beta$ is fixed at 0.59.